\documentclass[twocolumn,showpacs,preprintnumbers,amsmath,amssymb]{revtex4}

\usepackage{graphicx}
\usepackage{dcolumn}
\usepackage{bm}

\begin{document}


\title{Temperature dependence of the impurity-induced resonant state
in Zn-doped Bi$_2$Sr$_2$CaCu$_2$O$_{8+\delta}$ by
Scanning Tunneling Spectroscopy}

\author{H. Kambara}
\author{Y. Niimi}
\author{M. Ishikado}
\author{S. Uchida}
\author{Hiroshi Fukuyama}

\affiliation{Department of Physics, Graduate School of Science,
The University of Tokyo, 7-3-1 Hongo, Bunkyo-ku, Tokyo 113-0033, Japan}

\date{\today}

\begin{abstract}
We report on the temperature dependence of the impurity-induced resonant
state in Zn-doped Bi$_2$Sr$_2$CaCu$_2$O$_{8+\delta}$ by scanning
tunneling spectroscopy at 30 mK$\le T \le$ 52 K.
It is known that a Zn impurity induces a sharp resonant peak in tunnel
spectrum at an energy close to the Fermi level.
We observed that the resonant peak survives up to 52 K.
The peak broadens with increasing temperature, which is
explained by the thermal effect.
This result provides information to understand the origin of
the resonant peak.
\end{abstract}


\maketitle

The scanning tunneling spectroscopy (STS) technique based on
scanning tunneling microscope (STM) enables us to measure
local electronic density of states (LDOS) in atomic scale.
So far, there have been a lot of important studies to investigate
the key mechanism of high $T_c$ cuprates
by STS \cite{renner1,renner2,yazdani,pan,hudson,hoffman,lang,lee}.
Pan {\it et al.} \cite{pan} reported STS imaging of the LDOS around
impurity sites at surface of Zn-doped Bi$_2$Sr$_2$CaCu$_2$O$_{8+\delta}$
(Bi2212).
On the Zn site, they observed a tunnel spectrum with a sharp peak
at the energy ($-1.5$ meV) slightly below the Fermi level ($E_F$) and
cross-shaped fourfold quasiparticle spatial distributions.
The origin of this near-zero-energy peak (NZEP) is usually considered as
the impurity-scattering resonant state \cite{salkola,kruis,martin,balatsky}
because Zn-impurity has a strong scattering potential \cite{fukuzumi}.

Salkola and co-workers \cite{salkola} considered quasiparticle scattering
from a repulsive $\delta$-potential impurity using the $T$-matrix approach
and derived the resonant state within the superconducting gap.
But it is not straightforward to explain why the LDOS on the
impurity site is the largest since the Zn impurity site is a strong
scattering center.
Martin {\it et al.} \cite{martin} calculated tunneling matrix elements
between an STM tip and the Cu-O plane.
They claimed that tunneling electrons are ``filtered'' and consequently
the largest intensity is measured on the Zn site.
Recently Tang and Flatt\'e \cite{tang} gave a more quantitative
explanation for the experimental results \cite {pan} by considering
spatially extended Zn-impurity potentials on the Cu-O plane.

On the other hand, another competitive interpretation based on the
Kondo effect exists. The Kondo resonance scenario \cite{polkov,sachdev}
arose after the NMR experiments \cite{mahajan,julien,bobroff,ouazi}
which showed that the four nearest neighbor Cu atoms surrounding a Zn
impurity possess local moments.
These polarized spins will form the spin-singlet state with quasiparticles.
Note that it is not the standard Kondo effect since the density of states
of the quasiparticles vanishes at the Fermi energy due to the $d$-wave
superconducting gap structure.
In this scenario, the strongest LDOS peak at Zn site can be naturally
explained without considering the filter effect.
However, the scenario assumes unrealistically weak potential scattering,
which is not consistent with the transport measurement \cite{fukuzumi}.
The origin of the NZEP is still a matter of debate.

To test these scenarios, measuring the temperature evolution of the NZEP
should be one of the key experiments.
If the Kondo resonance scenario is correct, the peak weight of the NZEP
will increase at $T < T_K$.
The value of $T_K$ is estimated about 15 K \cite{polkov}
from the measured peak energy of $-1.5$ meV \cite{pan}.
In this paper, we report on the temperature dependence of NZEP in
Zn-doped Bi2212 in a temperature range from 30 mK to 52 K
using the ultrahigh vacuum (UHV) compatible STM \cite{ult-stm}.

The samples are Bi$_2$Sr$_2$Ca(Cu$_{1-x}$Zn$_x$)$_2$O$_{8+\delta}$
single crystals grown by the floating zone method (the superconducting
transition temperature $T_c = 89$ K, nominal $x = 0.6$\%).
They are cleaved at 100 K below $1\times10^{-7}$ Pa and then cooled to 30 mK.
The data were taken during subsequent warming up to 52 K.
We used an electrochemically etched tungsten wire for STM tips.
The STS measurements were performed by lock-in technique
with a modulation amplitude of 0.50 mV$_{\rm rms}$ and a
frequency of 411.7 or 511.7 Hz.
It takes typically 24 hours to obtain an STS image of 128$\times$128 pixels.
It was crucial to keep the temperature variations within $\pm1$\%
at $T \ge 20$ K to avoid unexpected tip crushes and thermal drifts
of the STS data.

\begin{figure}[t]
\includegraphics[width=1.0\linewidth]{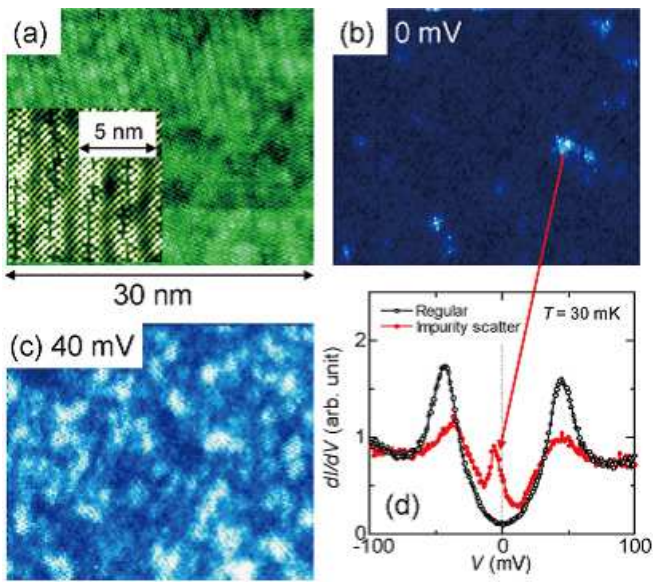}
\caption{\label{sts30mk}
(Color online)
STS data of Zn-doped Bi2212 at $T=30$ mK.
(a) Topographic image ($V=-0.12$ V, $I=0.2$ nA, $30\times25$ nm$^2$).
The inset shows a magnified topographic image
($V=-0.20$ V, $I=0.2$ nA, $10\times10$ nm$^2$).
(b) $dI/dV$ image at $V=0$ mV ($30\times25$ nm$^2$) of the same area
as that for the main topographic image in (a).
The bright spots denote the near-zero-energy peaks caused by
the Zn impurities.
(c) $dI/dV$ image at $V=40$ mV ($30\times25$ nm$^2$).
The bright patches roughly correspond to the regions of
the superconducting coherence peak of $\sim40$ meV.
(d) Typical tunnel spectra at a Zn impurity site and
away from impurity sites, respectively.}
\end{figure}

In Fig.~\ref{sts30mk}(a)-(d), we show the STS data of the cleaved
surface of Zn-doped Bi2212 obtained at $T=30$ mK.
Figure \ref{sts30mk}(a) shows a topographic image.
The inset is a magnified topographic image but on a different
surface where the atomic corrugation and supermodulation of the BiO
layer \cite{kirk} are more clearly seen.
Figure \ref{sts30mk}(b) is a $dI/dV$ image at a bias voltage $V=0$ mV
of the same area as that for the main topographic image
in Fig.~\ref{sts30mk}(a).
The NZEPs are visible as several bright spots here.
The apparent number density of the NZEP spots within this area
is about 0.2\% which is in the same order as $x=0.6$\%,
a nominal doping concentration of Zn.
Figure \ref{sts30mk}(c) is a $dI/dV$ image at $V=40$ mV.
The patch structure of the contrast indicates the inhomogeneous
distribution of the superconducting gap structure \cite{lang}.
In the bright patches of a few nm wide, the superconducting coherence
peak energy would be around 40 meV.
In Fig.~\ref{sts30mk}(d), we show a typical tunnel spectrum
at a Zn impurity site with the NZEP (solid circle) and the
superconducting gap structure obtained far away from the impurities
without the NZEP (open circle).
The NZEP spectra are characterized by depressed superconducting
coherence peaks as well as by sharp peaks near the Fermi energy.

\begin{figure}[t]
\includegraphics[width=1.0\linewidth]{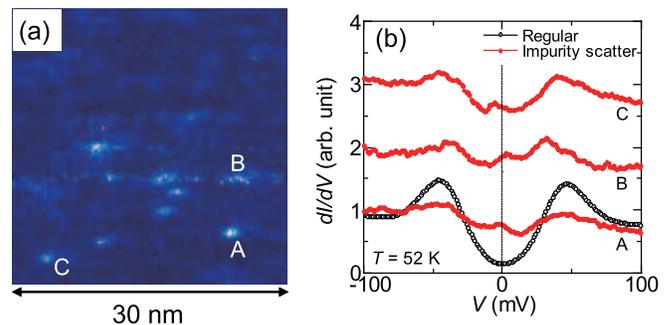}
\caption{\label{sts52k}
(Color online)
STS data measured at $T=52$ K at the different surface position from
Fig.~\ref{sts30mk}.
(a) $dI/dV$ image at $V=0$ mV ($30\times30$ nm$^2$).
The tunneling parameter for stabilization is the same
as Fig.~\ref{sts30mk} ($V=-0.12$ V, $I=0.2$ nA).
(b) Tunnel spectra on the three different Zn impurity sites A-C
denoted in (a) (solid circles) and that with a regular superconducting
gap (open circle) away from the impurity sites.
The spectra A-C are offset by 1 unit for clarity.}
\end{figure}

At a higher temperature of 52 K (Fig.~\ref{sts52k}), we obtained
a similar STS image with several bright spots to that in
Fig.~\ref{sts30mk}(b).
This means that the NZEP survives even at this temperature.
It is consistent with the fact that the broadened NZEPs
are observed in the tunnel spectra at three different sites
(Fig.~\ref{sts52k}(b)).
Note that the scan area of Fig.~\ref{sts52k}(a) is different from
that of Fig.~\ref{sts30mk}(b).

\begin{figure}[t]
\includegraphics[width=1.0\linewidth]{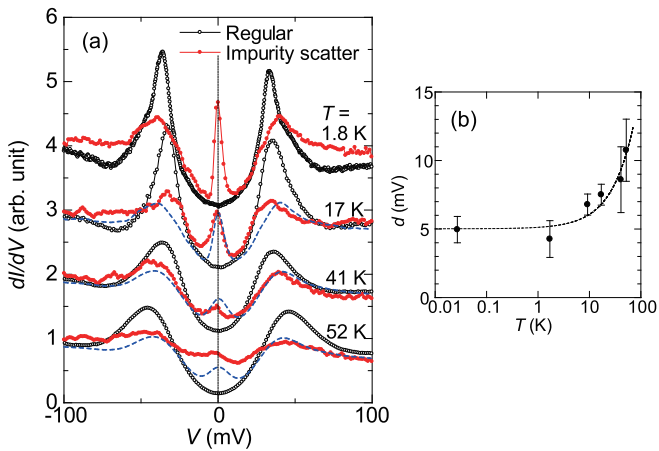}
\caption{\label{tdep}
(Color online)
(a) Temperature dependence of the tunnel spectra.
Typical superconducting gap structures (open circles) and
near-zero-energy peaks (solid circles) obtained at each temperature.
The dashed lines are calculated spectra based on the 1.8-K data
taking account of the thermal broadening effect in the Fermi
distribution function.
Each curve is shifted by 1 unit for clarity.
(b) Temperature dependence of full-width at half maximum ($d$)
of near-zero-energy peak.
The dashed line is a linear fit ($d$(mV)$=5.0+1.2k_{\rm B}T$).
Each error bar represents a standard deviation.}
\end{figure}

The temperature dependence of tunnel spectra is summarized
in Fig.~\ref{tdep}(a).
The NZEP smears out, {\it i.e.}, decreasing peak height, and increasing
peak width with increasing temperature up to 52 K.
The peak energy is determined as $-0.8 \pm 2.4$ meV at all the
temperatures we studied.
It scatters fairly largely impurity to impurity presumably due to
different scattering potentials.
It is difficult to determine the peak height precisely
since we averaged the spectra over $6-18$ positions
around the impurity although it is very sensitive to the exact
tip-location with respect to the impurity site.
These curves are not obtained on the same Zn impurity atom.
Nevertheless, we emphasize that the NZEPs certainly exist even at
high temperatures \cite{matsuda}. The decrease of the NZEP height
with increasing temperature seems to be well explained by the thermal
broadening effect as seen in Fig.~\ref{tdep}(a).
Here the dashed lines are calculated spectra based on the 1.8-K data
taking account of the thermal broadening effect in the Fermi
distribution function.

Figure \ref{tdep}(b) shows the temperature dependence of full-width
at half maximum of NZEP ($d$). The peak width is insensitive to the
averaging around the same impurity.
The increase of $d$ at higher temperatures above 20 K shows the thermal
broadening effect since the dashed line ($d$(mV)$=5.0+1.2k_{\rm B}T$)
represents the experimental data fairly well.
The intrinsic width ($d_0=5.0$ mV) below 2 K is similar to that
obtained by Pan {\it et al.} \cite{pan}.

We carried out STS measurement in a magnetic field of 6 T at $T=2$ K.
We observed the similar width ($\sim 6$ mV) of NZEP to that
obtained in zero field.
Pan {\it et al.} also reported no significant field dependence
of $d_0$ between 0 and 7 T \cite{pan2}.
The Zeeman splitting energy should be about 0.4 meV at
$B=6$ T \cite{grimaldi}. This is hard to be detected in our
measurement since it is much smaller than $d_0$.
We also note that the field of 6 T is too low to break up
the Kondo singlet \cite{vojta}.

Let us now discuss the origin of the NZEP.
At first, according to the Kondo resonance scenario,
it is predicted that the broadened Kondo peak still survives
almost up to $T_c$ in spite of $T_K < T_c$ \cite{vojtabulla}.
The temperature dependence of the peak weight becomes weaker above $T_K$.
Thus, only the existence of the NZEP at $T > T_K$ does not mean
straightforwardly the relevance of the Kondo resonance scenario.
Next, let us consider the impurity-scattering resonance scenario further
from the viewpoint whether superconductivity is crucial or not.
According to the calculation by Kruis {\it et al.} \cite{kruis}
and Balatsky {\it et al.} \cite{balatsky}, superconductivity is
unnecessary to form the impurity-induced resonant state.
Thus, they claim that the NZEP will be observed even in the
pseudo-gap region ($T > T_c$).
On the other hand, there are arguments that the Andreev resonance
in unconventional superconductor is the origin of the NZEP.
The quasiparticle scattering with sign change of the pair potential
results in the resonant state \cite{tanaka,tsuchiura,kashiwaya}.
This is observed in the tunnel junction experiments, for example,
by Iguchi {\it et al.} \cite{iguchi}. The multiple Andreev scattering
around a surface impurity will form the NZEP.
Future STS experiments for the pseudo-gap phase in the under-doped regime
would discriminate these two possibilities.

%
In summary, we measured the temperature dependence of near-zero-energy peak
(NZEP) in Zn-doped Bi$_2$Sr$_2$CaCu$_2$O$_{8+\delta}$.
The NZEPs are clearly observed up to 52 K with thermal broadening.
This result provides an important hint to understand the origin of the NZEP.

%
%
This work was financially supported by Grant-in-Aid for Specially
Promoted Research (No.~10102003) and Grant-in-Aid for Scientific
Research on Priority Areas (Nos.~17071002,~17071007) from MEXT, Japan,
and Grant-in-Aid for Young Scientists (A) from JSPS (No.~17684011).
Y.N. acknowledge the JSPS Research program for Young Scientists.
%
%
%

%
%
%
\end{document}